\newcommand{\CLASSINPUTtoptextmargin}{.75in}
\newcommand{\CLASSINPUTbottomtextmargin}{1in}
\newcommand{\CLASSINPUTinnersidemargin}{0.625in}
\documentclass[conference]{IEEEtran}
\makeatletter
\def\ps@headings{%
\def\@oddhead{\mbox{}\scriptsize\rightmark \hfil \thepage}%
\def\@evenhead{\scriptsize\thepage \hfil \leftmark\mbox{}}%
\def\@oddfoot{}%
\def\@evenfoot{}}
\makeatother
\IEEEoverridecommandlockouts

\usepackage{physics}
\usepackage{pifont}
\usepackage{amssymb, amsmath,amsfonts,bm,bbm,amsthm}
\usepackage[capitalise]{cleveref}
\usepackage{enumitem}

\newtheorem{theorem}{Theorem}

\newcommand{\opn}[1]{\operatorname{#1}}

\newcommand{\aspas}[1]{``#1''}
\renewcommand{\vu}[1]{\ensuremath{\hat{#1}}}

\allowdisplaybreaks
\newcommand{\opa}{\vu{a}}
\newcommand{\opad}{\vu{a}^{\dagger}}
\newcommand{\opadt}{\vu{a}^{\dagger T}}
\newcommand{\opb}{\vu{b}}
\newcommand{\opbd}{\vu{b}^{\dagger}}

\newcommand{\opq}{\vu{q}}

\newcommand{\opp}{\vu{p}}

\newcommand{\opr}{\vu{r}}

\usepackage{subcaption}

\usepackage{pgf}
\usepackage{pgfplots}
\usepackage{pgfplotstable}
\long\def\beginpgfgraphicnamed#1#2\endpgfgraphicnamed{\includegraphics{#1}}

\usepackage{tikz}
\pgfrealjobname{main}
\usetikzlibrary{plotmarks} 
\pgfplotsset{compat=1.6}

\usetikzlibrary{positioning,spy,shapes,shapes.misc,arrows,calc,decorations.markings}
\usetikzlibrary{patterns}
\usetikzlibrary{automata}

\pgfplotsset{graf/.style= {
	width=\columnwidth, height = 0.6\columnwidth,
    legend pos = north west,
    no markers, 
    xtick distance = {0.2},
    ymin=1.5, ymax = 2.5, 
    xmin=0, xmax = 1,
    ylabel= bits,
    xlabel= $\tau$}}

\tikzset{phase/.style={circle, draw, thick, anchor=center, fill = gray!20, minimum size = 0.4cm}}

\tikzset{detec/.style = {semicircle,draw,fill=blue!20,shape border rotate=270, minimum size=0.5cm, anchor=chord center}}

\tikzset{ahead/.style = {>=triangle 45, ->}}

\tikzset{->-/.style={>=triangle 45, decoration={
			markings,
			mark=at position 0.5 with {\arrow{>}}},postaction={decorate}}}

\tikzset{nd11/.style = {rectangle,anchor = center, draw, thick, minimum height = 0.9cm, minimum width = 3cm, rounded corners=.2cm}}

\tikzset{nd22/.style = {rectangle,anchor = center, draw, thick, minimum height = 1.8cm, minimum width = 3cm, rounded corners=.2cm}}

\tikzset{nd2/.style = {rectangle,anchor = center, draw, thick, fill = gray!30, minimum height = 1.5cm, minimum width = 1cm, rounded corners=.2cm}}

\tikzset{ahead2/.style = {>=triangle 45, <-}}

\tikzset{meter/.append style={draw, inner sep=10, rectangle, font=\vphantom{A}, minimum width=30, line width=.8,
 path picture={\draw[black] ([shift={(.1,.3)}]path picture bounding box.south west) to[bend left=50] ([shift={(-.1,.3)}]path picture bounding box.south east);\draw[black,-latex] ([shift={(0,.1)}]path picture bounding box.south) -- ([shift={(.3,-.1)}]path picture bounding box.north);}}}

\def\displacement#1#2#3{
	\begin{scope}[shift={#1}]
		\node[rectangle, draw, anchor=center, thick, fill = gray!20, minimum width = 1cm, minimum height = 0.4cm, rounded corners=.2cm] () at (0,0) {#2};		
		\node[inner sep=0cm] (#3i) at (-0.5,0) {}; 
		\node[inner sep=0cm] (#3o) at (0.5,0) {}; 		
\end{scope}}

\begin{document}

    \title{Evaluating the Eavesdropper Entropy via Bloch-Messiah Decomposition}
    
    \author{\IEEEauthorblockN{1\textsuperscript{st} Micael A. Dias}
    \IEEEauthorblockA{\textit{Department of Electrical Engineering} \\
    \textit{Federal University of Campina Grande}\\
    Campina Grande, Brazil\\
    micael.souza@ee.ufcg.edu.br}
    \and
    \IEEEauthorblockN{2\textsuperscript{nd} Francisco M. de Assis}
    \IEEEauthorblockA{\textit{Department of Electrical Engineering} \\
    \textit{Federal University of Campina Grande}\\
    Campina Grande, Brazil\\
    fmarcos@dee.ufcg.edu.br}
    }
    
    \maketitle
    
    \begin{abstract}
        We explore the Bloch-Messiah decomposition of Gaussian unitary to analyze the Entangling Cloner Attack performed by an eavesdropper on a discrete modulated continuous variable QKD scenario. Such a decomposition allows to replace the nonlinear unitary resulting from eavesdropping and tracing out Bob's mode into an architecture of single-mode operations (squeezers, phase shifters and displacements) and a two-mode beam splitter. Based on such architecture we were able to get tighter upper  bounds to the eavesdropper entropy for a discrete modulated CVQKD scheme. The new bounds are justified from the Gaussian extremality property valid for entangled-based equivalent protocols.
    \end{abstract}
    
    \begin{IEEEkeywords}
        Bloch-Messiah decomposition, CVQKD, Discrete Modulation.
    \end{IEEEkeywords}
    
    \section{Introduction}\label{sec:intro}
        
    In a Quantum Key Distribution (QKD) protocol, two legitimate parties (Alice and Bob) use a quantum channel to transmit random classical information and perform the task of distilling a completely random and secure bit string to be used as a secret key in symmetric cryptography setups under the eminence of a powerful eavesdropper (Eve) controlling the quantum channel and trying to retrieve information from the key \cite{pirandola2020}. The security of such protocols relies mainly on two fundamental concepts of quantum mechanics, the no-cloning theorem and the uncertainty principle \cite{nielsen2010}, being the only constrains to which the eavesdropper is submitted to and one must assume that she is able to perform any physically limited attack strategy. 

    In general, Alice and Bob exchange quantum states to generate correlated random sequences and will continue to post processing it by using a classical authenticated communication channel. On the other hand, the eavesdropper will deploy some attack strategy during quantum state distribution as an attempt to retrieve information by coupling ancillas to the states sent by Alice and measuring them. Each QKD protocol will then have different security degrees depending on assumptions on the eavesdropper attack capabilities \cite{cerf2007,weedbrook2012}.

    Most CVQKD protocols security analysis assume a collective Gaussian attack \cite{laudenbach2018}, which is not the most powerful attack possibly performed but it is quite strong: Eve couples an ancilla with each state sent by Alice by using an entangling cloner and performs a collective delayed measurement. 
    The entangling cloner provides the Gaussian character to the coupling model (and hence, a Gaussian channel \cite{pirandola2008}), which ensures maximum information to the eavesdropper as a result of the gaussian extremality theorem (GET) \cite{garcia-patron2006,wolf2006}. This is then a suitable scenario for Gaussian Modulated Coherent State (GMCS) protocols security analysis: Gaussian modulation in the Prepare and Measure (P\&M) protocol resulting on a Gaussian ensemble can be replaced by EPR states in the Entangled Based (EB) equivalent protocol and is reasonable to assume the channel output to be also Gaussian. This allows to compute the entropic quantities from the bipartite state covariance matrix. 

    However, protocols with non-Gaussian modulation brings new cards to the table. Some security proofs for these protocols remain relying on the GET, meaning that even when Alice and Bob certainly know that they do not used a Gaussian ensemble, they \textit{assume} it is Gaussian as it gives an upper bound on Eve's knowledge \cite{leverrier2009,zhao2020}. Of course it is a conservative option but a valid question one may rise is how much the eavesdropper information is being overestimated when non-Gaussian ensembles are treated as Gaussian. Providing more accurate methods for bounding this quantity has a direct relation to the expected performance of discrete modulated CVQKD protocols. 

    In order to approach this question, one can not call on the protocol's EB version as it can not be completely described by the first and second moments of a purified bipartite state. The analysis should follow the P\&M protocol and computing the eavesdropper entropic quantities depends on knowing how her state looks like after the entangling cloner, which, by tracing out Bob's mode, may be seen as an EPR state $\ket{\nu}$ undergoing some unitary $\vu{U}_i$ conditioned to the state $\vu\rho_i$ sent by Alice. Then, the ensemble of coherent states sent by Alice results on a non-Gaussian average state $\vu\rho_{Eve} = \sum_ip_i\vu{U}_i\op{\nu}\vu{U}_i^\dagger$ on Eve's modes with $p_i$ being the probability that $\vu\rho_i$ was sent.

    In this paper we attend to propose a method for computing the post entangling cloner eavesdropper ensemble entropy by applying the Bloch-Messiah (BM) decomposition on $\vu{U}_i\op{\nu}\vu{U}_i^\dagger$. This BM decomposition allows to express complex non linear evolutions as combinations of fundamental unitary and, combined with thermal decomposition, we find that $\vu\rho_{Eve}$ has the same entropy of an ensemble of displaced thermal states whose entropy can be computed using either the GET or the ensemble's Gramm matrix.

    The paper is structured as follows. In \Cref{sec:preliminary} we review some concepts of Gaussian systems, as the unitary general description and the so-called \textit{fundamental unitary}, and defines the BM recomposition. \Cref{sec:the_entangling_cloner_attack} explain the entangling clonner and in \Cref{sec:decomposing_the_eavesdropper_state} we apply the BM decomposition on the eavesdropper state. \Cref{sec:the_eavesdropper_entropy} shows how to compute the entropy from the decomposed states and in \Cref{sec:conclusions} we give our considerations and perspectives.

	\subsection{Notation}

	In the following, we denote linear operators with upper case letters, $\vu{D}$, and density operators with Greek low case letters, $\vu\rho$. Matrices and vectors comes as bold upper and lower case, $\bm{M}$ and $\bm{x}$, respectively, and we reserve $\bm{I}$, $\bm{X}$, $\bm{Y}$ and $\bm{Z}$ to be the Pauli matrices. The Hermitian conjugate of $\bm{M}$ is given by $\bm{M}^\dagger = (\bm{M}^*)^T$, the transpose conjugate. The canonical bosonic operator for the $i$-th mode comes as $\opa_i$ and, in the vectorial form, $\bm\opa = (\opa_1, \cdots, \opa_n)^T$ for a $n$-mode system. We take the quadrature operators to be in SI units, $\opq_i = \opa_i+\opad_i$ and $\opp_i = -i(\opa_i-\opad_i)$, and the vector of operators $\bm\opr = (\opq_1, \opp_1, \cdots, \opq_n, \opp_n)^T$.
    \section{Preliminary} 
\label{sec:preliminary}

    \subsection{Gaussian Unitary}
    
        Quantum operations model the quantum state evolution as a linear map $\mathcal{E}: \vu\rho \rightarrow \mathcal{E}(\vu\rho)$, which is completely positive and in the case of trace preserving ($\tr(\mathcal{E}(\vu\rho)) = 1$) it is also called a quantum channel. When a quantum channel is reversible, it is represented by a unitary transformation $\vu{U}$, $\vu{U}^{-1} = \vu{U}^T$. Then, within this scope, we say that a completely positive trace preserving reversible quantum operation is Gaussian when it transforms Gaussian states into Gaussian states. Such unitary are generated via a Hamiltonian $\vu{H}$ which are second order polynomials on the canonical operators, $\vu{U} = \exp{-i\vu{H}/2}$ and have the general form         
        \begin{equation}
        	\vu{H} = i(\bm\opad\bm\alpha+\bm\opadt\bm{A}\bm\opa+\bm\opadt\bm{B}\bm\opad)+\operatorname{H.c.},
        \end{equation}
        
        \noindent where $\bm\alpha\in\mathbb{C}^N$, $\bm{\opa} = (\opa_1, \cdots, \opa_N)^T$ is the vector of anihilation operators, $\bm{A}$ and $\bm{B}$ are $N\times N$ complex symmetric matrices and H.c. stands for the Hermitian conjugate. Such a unitary corresponds to the following Bogoliubov transformation in the Heisenberg picture        
        \begin{equation}
        	\bm\opa \rightarrow \bm{\opb =} \vu{U}^\dagger\bm\opa\vu{U} = \bm{E}\bm\opa + \bm{F}\bm\opad+\bm\alpha,
        \end{equation}
        
        \noindent being $\bm{E}$ and $\bm{F}$ complex matrices satisfying the constrains $\bm{EF}^T = \bm{FE}^T$ and $\bm{EE}^\dagger = \bm{FF}^\dagger+\bm{I}$, called the Bogoliubov matrices, and $\bm{\opb}$ the vector of anihilation operators on the output field. The unitary evolution of both creation and anihliation operators in the Heisenberg pucture may be arranged in the following block matrix form        
        \begin{equation}
        	\begin{pmatrix} \bm\opa \\ \bm\opad	\end{pmatrix} \rightarrow \begin{pmatrix} \bm\opb \\ \bm\opbd \end{pmatrix} = \begin{pmatrix} \bm{E} & \bm{F} \\ \bm{F}^* & \bm{E}^*	\end{pmatrix}\begin{pmatrix} \bm\opa \\ \bm\opad \end{pmatrix} + \begin{pmatrix}\bm{\alpha} \\ \bm{\alpha}^* \end{pmatrix}.
        \end{equation}
        
        Analougsly to the Bogouliubov transformation, which relates the input and output canonical field operators, we may define a more simple description of Gaussian unitary through the evolution of quadrature operators by an affine map         
        \begin{equation}
        	\bm{\vu{r}} \rightarrow \bm{S}\bm{\vu{r}} + \bm{d},
        \end{equation}        
        \noindent where $\bm{\vu{r}} = (\opq_1,\opp_1,\cdots, \opq_N,\opp_N)^T$, being $\opq_i = \opa_i+\opad_i$ and $\opp_i = -i(\opa_i - \opad_i)$ the corresponding position and momenta field operators for the $i$-th mode, $\bm{S}$ is a $2N\times 2N$ real symplectic matrix and $\bm{d}\in\mathbb{R}^{2N}$. Given the direct relation between canonical bosonic operators and the position and momentum operators, it is possible to retrieve $\bm{S}$ if $\bm{E}$ and $\bm{F}$ are given, and \textit{vise versa}.
        

        
        
        
    
	\subsection{Fundamental Unitary}
	
		We highlight three specific Gaussian unitary operations, namely, the Displacement, Squeezing and Rotation operators, which factorize any arbitrary Gaussian unitary.
		
		\begin{enumerate}[label=\roman*)]
			\item The $N$-mode displacement is given by the following operator			
			\begin{equation}
				\vu{D}_{\bm\alpha} = \exp(\bm\alpha^T\bm\opad-\bm\alpha^\dagger\bm\opa),
			\end{equation}			
			\noindent where $\bm\alpha = (\alpha_1,\cdots,\alpha_N)^T \in\mathbb{C}^N$ and $\alpha_i = q_i+ip_i$. The respective Bogoliubov matrices are $\bm{E} = \bm{I}$ and $\bm{F} = \bm{0}$ with displacement vector $\bm{\alpha}$ for a complete transformation expression. Moreover, the symplectic resulting on the quadrature operators is given by			
			\begin{align}
				\bm{\vu{r}} \rightarrow \bm{\vu{r}} + \bm{d}_\alpha, & & \bm{d}_\alpha = (q_1,p_1,\cdots,q_N,p_N)^T.
			\end{align} 
			
			\item The $N$-mode rotation operator is specified by the $N\times N$ hermitian matrix $\bm{\phi}$,			
			\begin{equation}
				\vu{R}_{\bm{\phi}} = \exp(i\bm{\opadt\phi\opa}),
			\end{equation}			
			\noindent corresponding to the Bogoliubov matrices $\bm{E} = e^{i\bm{\phi}}$ and $\bm{F} = \bm{0}$, with null displacement vector. 
			
			\item The general $N$-mode squeezing operator is defined by the $N\times N$ symmetric matrix $\bm{Z}$ 			
			\begin{equation}
				\vu{S}_{\bm{Z}} = \exp(\frac{1}{2}(\bm{\opadt Z\opad - \bm{\opa^TZ^\dagger\opa}})).
			\end{equation}
			
			The squeezing matrix $\bm{Z}$ may be polar decomposed as $\bm{Z} = \bm{r}e^{i\bm\theta}$. Then, the Bogoliubov matrices $\bm{E} = \cosh(\bm{r})$ and $\bm{F} = \sinh(\bm{r})e^{i\bm{\theta}}$ and null displacement vector.
		\end{enumerate}
    
	\subsection{Switching Rules}

		The fundamental unitary operators do not possess the convenience of commuting with each other, but, according to \cite{ma1990} they can be properly switched with proper parameter adjustments, called the \emph{switching rules}:		
		\begin{align}
			\vu{D}_{\bm\alpha}\vu{S}_{\bm{Z}} &= \vu{S}_{\bm{Z}}\vu{D}_{\bm\beta}, & \bm\beta &= \cosh(\bm{r})\bm\alpha - \sinh(\bm{r})e^{i\bm\theta}\bm\alpha^*,\\
			\vu{S}_{\bm{Z}}\vu{R}_{\bm{\phi}} &= \vu{R}_{\bm{\phi}}\vu{S}_{\bm{Z'}}, & \bm{Z'} &= e^{-i\bm\phi}\bm{Z}e^{-i\bm\phi^T},\\
			\vu{D}_{\bm\alpha}\vu{R}_{\bm{\phi}} &= \vu{R}_{\bm{\phi}}\vu{D}_{\bm\gamma}, & \bm\gamma &= e^{-i\bm\phi}\bm\alpha.
		\end{align}
    	
    \subsection{Bloch-Messiah Decomposition}
    
        The Bloch-Messiah (BM) decomposition uses a specific simultaneous and \aspas{conditioned} solution for a singular value decomposition (SVD) of Bogoliubov matrices $\bm{E}$ and $\bm{F}$ in order to split complicated non-linear Gaussian unitary into a sequence of rotation, squeezing and displacement operations \cite{braunstein2005,cariolaro2016,cariolaro2016a}. In the following, we revisit the main point of the BM decomposition.
        
        \begin{theorem}[Bloch-messiah Decomposition \cite{braunstein2005}]\label{th:bm-decomp}
        	For arbitrary Bogoliubov matrices $\bm{E}$ and $\bm{F}$ it is possible to find a specific decomposition assuming the form        	
        	\begin{align}\label{eq:bm1}
        		\bm{E} &= \bm{U\Lambda_EW}_E^\dagger, & \bm{F} &= \bm{U\Lambda_FW}_F^\dagger,
        	\end{align}        	
        	\noindent where $\bm{U}$, $\bm{W}_E$ and $\bm{W}_F$ are unitary matrices satisfying        	
        	\begin{equation}\label{eq:rotation-cond}
        		\bm{W}_F = \bm{W}^*_E,
        	\end{equation}        	
        	\noindent which is commonly called the \textit{rotation condition}, and $\bm{\Lambda}_E$ and $\bm{\Lambda}_F$ are diagonal with nonnegative entries that satisfy        	
        	\begin{equation}
        		\bm{\Lambda}_E = \bm{I} + \bm{\Lambda}_F.
        	\end{equation}
        \end{theorem}
        
        
        As stated before, the BM decomposition requires a very specific SVD on both Bogoliubov matrices representing the arbitrary Gaussian operation. First, the SVD must have the same unitary left matrix, which is possible once $\bm{E}$ and $\bm{F}$ are diagonal on the same basis. Second, and more subtle, \Cref{eq:rotation-cond} establishes that the SVD unitary matrices on the right are not arbitrary. In fact, this condition is not always satisfied for an arbitrary SVD solution and one must perform a two-step procedure: (i) perform the SVD that satisfies \Cref{eq:bm1}, which most of times does not satisfy the rotation condition, and (ii) from the matrices $\bm{W}_E$ and $\bm{W}_F$ obtained, evaluate the balancing matrix $\bm{D}$ from the Takagi factorization as defined bellow \cite{cariolaro2016}.
        
        \begin{theorem}[Takagi Factorization {\cite[Corollary 4.4.4]{horn2012}}]\label{th:takagi}
        	A complex symmetric $N\times N$ matrix $\bm{A}$ can be decomposed in the form         	
        	\begin{equation}
        		\bm{A} = \bm{U}_A\bm\Lambda_A\bm{U}_A^T,
        	\end{equation}        	
        	\noindent where $\bm{U}_A$ is a unitary matrix and $\bm\Lambda_A$ is diagonal with non-negative entries, the singular values of $\bm{A}$. Particularly, if $\bm{A}$ is symmetric and unitary,         	
        	\begin{equation}
        		\bm{A} = \bm{U}_A\bm{U}_A^T \longrightarrow \bm{U}_A = \bm{A}^{1/2}.
        	\end{equation}
        \end{theorem}
        
        
        Then, from the matrices $\bm{W}_E$ and $\bm{W}_F$ we compute the matrix $\bm{G} = \bm{W}_E^\dagger\bm{W}^*_F$ which is block diagonal, unitary and symmetric and, according to the Takagi factorization, $\bm{G} = \bm{DD}^T$. Then, we can conclude the Bloch-Messiah decomposition by introducing the balancing matrix $\bm{D}$ in the previous unitary matrices as $\bm{\mathcal{U}} = \bm{UD}$, $\bm{\mathcal{W}}_E = \bm{W}^*_F\bm{D}^*$ and $\bm{\mathcal{W}}_F = \bm{W}_F\bm{D}$, which results in	        
        \begin{align}
        	\bm{E} = \bm{\mathcal{U}}\bm\lambda_E\bm{\mathcal{W}}_E^\dagger, & & \bm{F} = \bm{\mathcal{U}}\bm\lambda_F\bm{\mathcal{W}}_F^\dagger.
        \end{align}

    \section{Discrete Modulated CVQKD and the Entangling Cloner Attack} 
\label{sec:the_entangling_cloner_attack}

	We begin by providing a quick overview of a Prepare \& Measure Continuous-Variable QKD Protocol (PMP) with discrete (non-Gaussian) modulation of coherent states. Define a set of complex amplitudes $\mathcal{A} = \qty{\alpha_1, \cdots, \alpha_K}$, for $K$ positive integer, and a discrete probability distribution $P = \qty{p_0, \cdots, p_K}$, which specifies the ensemble $\mathcal{S} = \qty{\ket{\alpha_i},p_i}$. Alice, then, prepares randomly states from $\mathcal{S}$ and send them to Bob through a quantum Gaussian channel. Bob performs either homodyne or heterodyne detection and, from his detection results, him and Alice will start the protocol's classical stage, performing parameter estimation, information reconciliation, and privacy amplification. 

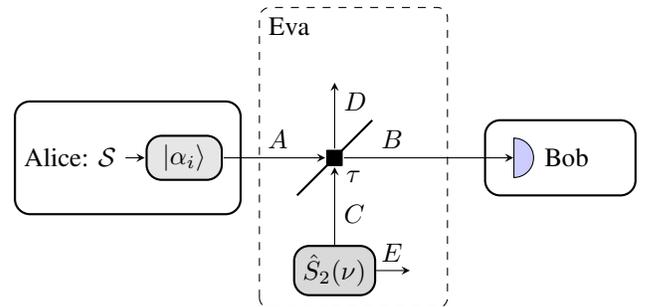
\begin{figure}[!b]
		\centering%
		\leavevmode\beginpgfgraphicnamed{ent-cln-atack}
		\begin{tikzpicture}[ >=stealth]
	\draw[draw, thick, rounded corners=.2cm] (-4.25,-.75) rectangle (-1.25,.75);
	\node[anchor=west] (a) at (-4.25,0) {Alice: $\mathcal{S}$};
	\displacement{(-2,0)}{$\ket{\alpha_i}$}{d1};
	\node (c) at (0,0) {};
	\draw[thick] (-.5,-.5) -- (.5,.5);
	\draw[fill=black] (-.1,-.1) rectangle (.1,.1);
	\node at (.25,-.25) {$\tau$};
	
	\draw[draw, thick, rounded corners=.2cm] (4,-.5) rectangle (2,.5);
	\node[semicircle,draw,fill=blue!20,shape border rotate=270, minimum size=0.25cm] (det) at (2.5,0) {};
	\node[anchor=east] at (3.5,0) {Bob};
	
	\node[draw, rectangle, rounded corners=.2cm, fill=gray!30, minimum height=2,thick] (s2) at(0,-1.5) {$\vu{S}_2(\nu)$};
	
	\node[anchor=south east] at (-.5,0) {$A$};
	\node[anchor=south west] at (.5,0) {$B$};
	\node[anchor=north west] at (0,-0.5) {$C$};
	\node[anchor=south west] at (0,0.5) {$D$};
	\node[anchor=south west] at (.5,-1.5) {$E$};
	
	\draw[->] (a) -- (d1i);
	\draw[->] (d1o) -- (c);
	\draw[->] (c) -- (det);
	\draw[->] (s2) -- (c);
	\draw[->] (c) -- (0,1);
	\draw[->] (s2) -- (1,-1.5);
	
	\draw[dashed,rounded corners] (-1,-2) rectangle (1.5,2);
	\node[anchor=north west] at (-1,2) {Eva};
\end{tikzpicture}
		\endpgfgraphicnamed
		\caption{\label{fig:eca}General scheme of an Entangling Cloner Attack (ECA).}
	\end{figure}

	The eavesdropper, on the other hand, will perform a physical attack simulating a non-eavesdropped thermal loss channel with transmittance $\tau$ and thermal noise $\varepsilon = 2\bar{n}+1$, where $\bar{n}$ is the mean number of thermal photons excited. In this physical attack, named the \textit{entangling cloner}, she replaces the thermal loss channel by a controlled beam splitter (BS) of transmittance $\tau_E$ and couples each state sent by Alice with one half of a TMSV state 
	\begin{equation}
		\ket{\nu}_{CE} = \sqrt{1-\lambda^2}\sum_{n=0}^{\infty}(-\lambda)^n\ket{n}_C\ket{n}_E,
	\end{equation}
	\noindent where $\lambda = \tanh(\frac{1}{2}\cosh^{-1}(\nu))$. Then, she sets $\tau_E = \tau$ and $\nu = \varepsilon = 2\bar{n}+1$ to match the thermal loss channel parameters without the presence of an eavesdropper. The BS output modes are Bob's ($B$) to measure as it is received and Eve's ($D$) to store in a quantum memory jointly with the second TMSV mode $E$ to perform a delayed collective measurement (collective attack strategy). The whole scheme is exemplified in \Cref{fig:eca}.


	We can represent these states and the beam spliter action in the symplectic phase space through its displacement vector and the covariance matrix as the states and operations are all Gaussian. Alice's and Eve's initial states are represented by the following covariance matrix
	\begin{equation}
	\bm\Sigma = \begin{pmatrix} \bm{I} & 0 & 0 \\ 0 & (2\Bar{n}+1)\bm{I} & 2\sqrt{\Bar{n}^2+\Bar{n}}\bm{I} \\ 0 & 2\sqrt{\Bar{n}^2+\Bar{n}}\bm{I} & (2\Bar{n}+1)\bm{I}
	\end{pmatrix},
	\end{equation}

	\noindent and the beam-splitting operation by the symplectic map
	\begin{equation}
		\bm{B}_T = \begin{pmatrix} t\bm{I}_2 & r\bm{I}_2 & 0 \\ -r\bm{I}_2 & t\bm{I}_2 & 0 \\ 0 & 0 & \bm{I}_2	\end{pmatrix},
	\end{equation}

	\noindent with $t = \sqrt{\tau}$ and $r = \sqrt{1-\tau}$. The final state after BS map is given by
	\begin{equation}
		\bm{B}_T\bm\Sigma\bm{B}_T^T = \begin{pmatrix} 
		(2r^2\Bar{n}+1)\bm{I}          & 2tr\Bar{n}\bm{I}_2               & 2r\sqrt{\Bar{n}^2+\Bar{n}}\bm{Z} \\ 
		2tr\Bar{n}\bm{I}_2               & (2t^2\Bar{n}+1)\bm{I}          & 2t\sqrt{\Bar{n}^2+\Bar{n}}\bm{Z} \\ 
		2r\sqrt{\Bar{n}^2+\Bar{n}}\bm{Z} & 2t\sqrt{\Bar{n}^2+\Bar{n}}\bm{Z} & (2\Bar{n}+1)\bm{I}
		\end{pmatrix},
	\end{equation}

	\noindent and, tracing out Bob's mode, one gets
	\begin{equation}\label{eq:eve-cov}
		\bm\Sigma_{Eve} = \begin{pmatrix}
		(2t^2\Bar{n}+1)\bm{I}          & 2t\sqrt{\Bar{n}^2+\Bar{n}}\bm{Z} \\ 
		2t\sqrt{\Bar{n}^2+\Bar{n}}\bm{Z} & (2\Bar{n}+1)\bm{I}
		\end{pmatrix} = \begin{pmatrix} a\bm{I} & c\bm{Z} \\ c\bm{Z} & b\bm{I}\end{pmatrix}.
	\end{equation}
		
	For the displacement vector, we have that Alice's coherent state and Eve TMSV state reads $\ev{\bm{\vu{r}}_i} = (q_i, p_i,0,0,0,0)^T$ and, $\bm{BS}_{tot}\ev{\bm{\vu{r}}_i} = (tq_i, tp_i, -rq_i, -rp_i,0,0)^T$. Tracing out Bob's mode, $\ev{\bm{\vu{r}}}_{eve|i} = (-rq_i, -rp_i,0,0)^T$.

    \section{Decomposing The Eavesdropper's state} 
\label{sec:decomposing_the_eavesdropper_state}

	Given the previous description of Eve's entangling cloner and how it changes her state after the coupling, we develop how the BM decomposition can be applied to her physical attack. Beginning with the covariance matrix of \cref{eq:eve-cov}, we have that it presents a standard matrix \cite{weedbrook2012} for which the symplectic eigenvalues are given by $\nu_{1,2} = [\sqrt{(a+b)^2-4c^2}\pm(b-a)]/2$ and the symplectic matrix for the thermal decomposition\footnote{The thermal decomposition follows from the Williamson's theorem for real positive semi-definite even dimensional matrices which states that any matrix satisfying these previous conditions can be put in diagonal form by a sympletic transformation. That is, for arbitrary symplectic $2N\times 2N$ matrix $\bm{V}$, one gets that $\bm{V} = \bm{S}\bm{V}^\oplus\bm{S}^T$, where $\bm{V}^\oplus = \opn{diag}(\nu_1,\nu_1,\cdots,\nu_N,\nu_N)$ and $\bm{S}$ is a symplectic map.} $\bm{\Sigma_{Eve}} = \bm{S}\bm{\Sigma_{Eve}}^\oplus\bm{S}^T$ is given by
	\begin{align}\label{eq:simp-eve}
	\bm{S} = \begin{pmatrix}
	w_1\bm{I} & w_2\bm{Z} \\ w_2\bm{Z} & w_1\bm{I}, 
	\end{pmatrix}, & & w_{1,2} = \sqrt{\frac{a+b}{2\sqrt{(a+b)^2-4c^2}}\pm\frac{1}{2}}.
	\end{align}

	From the symplectic matrix in \cref{eq:simp-eve} we can compute the Bogoliubov matrices using the relations between bosonic and quadrature operators. Then,
	\begin{align}
	\label{eq:eve-bog-matrices}	
	\begin{pmatrix}
	\bm\opb \\ \bm\opb^\dagger
	\end{pmatrix} &= \begin{pmatrix}
	w_1\bm{I} & w_2\bm{X} \\
	w_2\bm{X} & w_1\bm{I} \\
	\end{pmatrix}
	\begin{pmatrix}
	\bm\opa \\ \bm\opa^\dagger
	\end{pmatrix},
	\end{align}	

	\noindent where the matrices $\bm{E} = w_1\bm{I}$ and $\bm{F} = w_2\bm{X}$ are the corresponding Bogoliubov matrices.


	Then, from the Bogoliubov matrices obtained in \Cref{eq:eve-bog-matrices}, we apply the decomposition described in the previous section in order to obtain the eavesdropper's Bloch-Messiah architecture:
	\begin{enumerate}
		\item Singular values of $\bm{E}$ and $\bm{F}$:		
		\begin{align}
			\bm{EE}^\dagger &= w_1^2\bm{I},  & \bm{FF}^\dagger &= w_2^2\bm{I},
		\end{align}
		
		\noindent once $w_1, w_2\in\mathbb{R}$.
		
		\item Singular value decomposition of $\bm{E}$ and $\bm{F}$		
		\begin{align}
			\bm{E} &= \bm{I\Lambda}_E\bm{I} = \bm{U\Lambda}_E\bm{W}_E^\dagger, \\
			\bm{F} &= \bm{I\Lambda}_F\bm{X}= \bm{U\Lambda}_F\bm{W}_F^\dagger,
		\end{align}
		
		\noindent where $\bm\Lambda_E = \opn{diag}(w_1,w_1)$ and $\bm\Lambda_F = \opn{diag}(w_2,w_2)$ the diagonal matrices referencing the squeezing operation and finally, $\bm{W}_E^\dagger = \bm{I}$ and $\bm{W}_F^\dagger = \bm{X}$ are the right rotation matrices which does not match the rotation condition.
		
		\item Compute $\bm{G} = \bm{W}_E^\dagger\bm{W}^*_F$		
		\begin{equation}
			\bm{W}_E^\dagger\bm{W}^*_F = \bm{X}.
		\end{equation}
		
		\item Compute the balancing matrix using \Cref{th:takagi} (Takagi factorization), 		
		\begin{align}
			\bm{G} &= \bm{DD}^T \rightarrow \bm{D} = \bm{G}^\frac{1}{2} = \bm{X}^\frac{1}{2}\\
			\bm{D} &= \frac{1}{\sqrt{2}}\begin{pmatrix} e^{i\pi/4} & e^{-i\pi/4} \\ e^{-i\pi/4} & e^{i\pi/4}.
			\end{pmatrix}
		\end{align}
		
		\item Compute the left and right rotation matrices using the balancing matrix $\bm{D}$,		
		\begin{align}
			\bm{\mathcal{W}}_E^\dagger &= \bm{D}^T\bm{W}_F^T = \frac{1}{\sqrt{2}}\begin{pmatrix} e^{-i\pi/4} & e^{\pi/4} \\ e^{i\pi/4} & e^{-i\pi/4}\end{pmatrix},\\ 
			\bm{\mathcal{U}} &= \bm{U}\bm{D} = \frac{1}{\sqrt{2}}\begin{pmatrix} e^{i\pi/4} & e^{-i\pi/4} \\ e^{-i\pi/4} & e^{i\pi/4}
			\end{pmatrix}.
		\end{align}
		\noindent and $\bm{\mathcal{W}}_F = \bm{\mathcal{W}}^*_E$.
	\end{enumerate}


	With the appropriated matrices, we conclude the BM decomposition of Eve's unitary transformation $\vu{U}$ represented by the Bogoliubov matrices $\bm{E} = \bm{\mathcal{U}}\bm\Lambda_E\bm{\mathcal{W}}_E^\dagger$ and $\bm{F} = \bm{\mathcal{U}}\bm\Lambda_F\bm{\mathcal{W}}_F^\dagger$ corresponds to a rotation operation $\vu{R}_{\phi_1}$, with $e^{i\bm\phi_1} = \bm{\mathcal{W}}_E^\dagger$, a parallel set of one mode squeezers $\vu{S}_{\bm{r}}$ where $\cosh(\bm{r}) = \bm\Lambda_E$ and $\sinh(\bm{r})e^{i\bm\theta} = \bm\Lambda_F$, and a second rotation operator $\vu{R}_{\phi_2}$ with $e^{i\bm\phi_2} = \bm{\mathcal{U}}$, that is,
	\begin{equation}
		\vu{U} = \vu{R}_{\phi_2}\vu{S}_{\bm{r}}\vu{R}_{\phi_1}.
	\end{equation}

	Then, by including the displacement, the TMSV state may be seen as undergoing the following transformation:
	\begin{equation}\label{eq:eve-state}
		\op{\nu}_{CE} \longrightarrow \vu{D}_{\bm\beta_i}\vu{U}\qty[\vu\rho^{th}_{\nu_1'}\otimes\vu\rho^{th}_{\nu_2'}]\vu{U}^\dagger\vu{D}_{\bm\beta_i}^\dagger = \vu\rho_{Eve|i},
	\end{equation}

	\noindent where we call $\vu\rho^{th}_{\nu_i'}$ the thermal state with photon number $\nu_i'$ with $\nu_1' = (\nu_1-1)/2$ and $\nu_2' = (\nu_2-1)/2$ and $\bm\beta_i = \ev{\bm{\vu{r}}}_{eve}$. As the unitary $\vu{U}$ does not depend on the state sent by Alice but only on the parameters $\tau$ and $\nu$, one has that Eve gains information by the displacement on one mode of her TMSV state while add some thermal noise on Bob's mode.

    \section{The Eavesdropper Entropy} 
\label{sec:the_eavesdropper_entropy}

	In \Cref{sec:the_entangling_cloner_attack} we described the general structure for the entangling cloner pervormed by Eve and provided the symplectic picture for her state with Bob's mode traced out. In \Cref{sec:decomposing_the_eavesdropper_state} the BM decomposition was used together with thermal decomposition to conclude that the ECA results on an unitary that does not depends on the coherent state sent by Alice, acting on a two-mode thermal state, and the classical information of interest lies on the displacement. In this section we look forward to compute Eve's entropy using the results of the previous sections and compare it with the bounds given by an EB framework.

    Once Alice prepares states from the ensemble $\mathcal{S}$, Eve's average state after the channel is
    \begin{equation}\label{eq:eve-av-state}
        \vu\rho_{Eve} = \sum_ip_i\vu\rho_{Eve|i},
    \end{equation}
    \noindent where $\vu\rho_{Eve|i}$ is given by \Cref{eq:eve-state}. Within the context of quantum key distribution, the eavesdropper information is given by Holevo bound which gives the maximum mutual information between Eve and Alice (or Bob) resulting from an optimal measurement performed by Eve. This bound, in reverse reconciliation, relate to Bob's outcomes and is given by
    \begin{equation}\label{eq:holevo}
        \chi(B;E) = S(\vu\rho_{Eve}) - \int p(b)S(\vu\rho_{Eve|b})\dd{b},
    \end{equation}    
    \noindent where $S(\sigma) = -\tr(\sigma\log\sigma)$ is the von Neumann entropy, the integration may be on the real line, if Bob homodynes, or on the complex plane if he heterodynes, $\vu\rho_{Eve|b}$ is Eve's average state given Bob's outcome $b$. 

    We turn our attention to the first therm on the left side of \Cref{eq:holevo}, the entropy of Eve's average state. The expression can be simplified by using the switching rules on the operators $\vu{D}$ and $\vu{\mathcal{U}}$:
    \begin{equation}
    \vu{D}_{\beta_i}\vu{R}_{\phi_2}\vu{S}_{\bm{r}}\vu{R}_{\phi_1} = \vu{R}_{\phi_2}\vu{S}_{\bm{r}}\vu{R}_{\phi_1}\vu{D}_{\beta'_i},
    \end{equation}
    \noindent where 
    \begin{align}
        \bm\beta'_i &= e^{-i\bm\phi_1}\cosh(\bm{r})e^{-i\bm\phi_2}\bm\beta_i - e^{-i\bm\phi_1}\sinh(\bm{r})e^{i\bm\phi_2}\bm\beta^*_i\\
         &= \bm{\mathcal{W}}_E^T\bm\Lambda_E\bm{\mathcal{U}}^*\bm\beta_i - \bm{\mathcal{W}}_E^T\bm\Lambda_F\bm{\mathcal{U}}\bm\beta^*_i\\
         &= w_+\bm\beta_i - w_-\bm\sigma_X\bm\beta^*_i\\
         &= (-w_1r\alpha_i, w_2r\alpha_i^*)^T,
    \end{align}
    \noindent and the corresponding phase space displacement vector becomes
    \begin{equation}
        \bm{d} = \qty(-w_1rq_i,-w_1rp_i, w_2rq_i,-w_2rp_i)^T.
    \end{equation}

    Then, Eve's state after the ECA given by \Cref{eq:eve-state} can also be expressed as 
    \begin{equation}
        \vu\rho_{Eve|i} = \vu{U}\vu{D}_{\beta'_i}\qty(\vu\rho^{th}_c(\nu_1')\otimes\vu\rho^{th}_e(\nu_2'))\vu{D}_{\beta'_i}^\dagger\vu{U}^\dagger,
    \end{equation}
    \noindent which is a displaced two-mode thermal state under the action of the operation $\vu{U}$ and the average state state entropy reads
	\begin{align}
		S(\vu\rho_{Eve}) &= S\qty(\sum_ip_i\vu\rho_{Eve|i})\\
		 &= S\qty(\vu{U}\qty[\sum_ip_i\vu{D}_{\beta'_i}\vu\rho^{th}_{\nu_1'}\otimes\vu\rho^{th}_{\nu_2'}\vu{D}_{\beta'_i}^\dagger]\vu{U}\dagger)\\\label{eq:thermal-displaced-state}
		 &= S\qty(\sum_ip_i\vu{D}_{\bm\beta'_i}\vu\rho^{th}_{\nu_1'}\otimes\vu\rho^{th}_{\nu_2'}\vu{D}_{\bm\beta'_i}^\dagger\vu{U}^\dagger)\\
         &= S\qty(\vu\rho).
	\end{align}
    \noindent as the von Neumann is invariant under unitary operations. One conclusion is that Eve's average state has the same entropy of a set of two-mode thermal states with suitable displacements, $\vu\rho$. 

    Even with a simpler expression after taking the unitary $\vu{U}$ out, it remains a non-trivial problem as $\vu\rho$ is not Gaussian. Yet, we highlight two ways of computing it. The first one is to treat $\vu\rho$ as Gaussian and use the GET to upper bound its entropy, obtained by the symplectic eigenvalues of its covariance matrix $\Sigma$. It will be then an upper bound on her entropy.
    The second way to obtain $S(\vu\rho)$ is to compute the Gramm matrix $M$ for the set of displaced thermal states and then compute the entropy of $M$. We recall that, for an ensemble $\mathcal{E} = \qty{\ket{\psi}_k, p_k}$ of pure states on finite dimensional systems, the normalized Gramm matrix $M$ with elements $\qty[M]_{m,n} = \sqrt{p_mp_n}\braket{\psi_m}{\psi_n}$ has the property of having the same entropy of $\mathcal{E}$, that is, $S(\mathcal{E}) = S(M)$ \cite{jozsa2000,hughston1993}. In the case of multimode Gaussian states, one oly needs to replace the overlap $\braket{\psi_m}{\psi_n}$ by the Hilbert-Schmidt product of Gaussian states. Although, this solution is still a conjecture as it is not proved that this Gramm matrix property on the entropy is still valid when the states are on infinite dimensional Hilbert spaces.

    Now, we shall exemplify how the above results provide a tighter bound (or an exact measure with the Gramm matrix conjecture) on Eve's entropy then the ones obtained by using an EB version of a discrete modulated CVQKD protocol. 
    Lets take as an example a QPSK constellation based CVQKD protocol \cite{leverrier2009}. In such a protocol, the P\&M version consists on Alice preparing coherent states equiprobably from the set $\qty{\ket{\alpha_1}, \ket{\alpha_2}, \ket{\alpha_3}, \ket{\alpha_4}}$ where $\alpha_k = \alpha e^{i\theta_k}$, $\theta_k = (2k-1)\pi/4$, and sending them through a thermal-loss quantum channel with transmittance $\tau$ and thermal noise $\Bar{n}$ and Bob will perform heterodyne detection at the reception. This P\&M protocol has an Entangled Based equivalent which is obtained by a proper purification of Alice's average state (a pure bipartite state $\ket{\Psi_4}$) and, by applying the GET, one can assume that Eve's entropy equals the bipartite state entropy. This entropy is obtained by the covariance matrix simplectic eigenvalues. 
    
\begin{figure}[!tb]
    	\centering
    	\begin{tikzpicture}
    		\pgfplotstableread{res/BM-EveEntr.txt}\BM
    		\pgfplotstableread{res/EveEntropy.txt}\GE
    		\pgfplotstableread{res/GM-Eve-Entr.txt}\GM
    		
    		\begin{axis}[graf]
    			
    			\addplot[red] table[x=0,y=2] {\BM};
    			\addplot[black] table[x=0,y=3] {\BM};
    			
    			\addplot[red, dashed] table[x=0,y=2] {\GE};
    			\addplot[black, dashed] table[x=0,y=3] {\GE};
    			
    			\addplot[red] table[x=0,y=1] {\GM};
    			\addplot[black] table[x=0,y=2] {\GM};
    			
    			\node[anchor=east, font=\scriptsize] () at (axis cs:1,2.18) {EB-QPSK.};
    			\node[anchor=east, font=\scriptsize] () at (axis cs:1,1.97) {BM-GET};
    			\node[anchor=east, font=\scriptsize] () at (axis cs:1,1.68) {BM-GME};
    			
    			\legend{$\bar{n} = 0.01$, $\bar{n} = 0.02$}
    		\end{axis}
    	\end{tikzpicture}
    	\caption{\label{fig:entropy}. Eve's entropy in a QPSK discrete modulation CVQKD scenario evaluated from the equivalent Entangled Based protocol (EB-QPSK) and the BM decomposition using either the Gaussian extremality theorem (BM-GET) and the Gramm matrix entropy (BM-GME).}
    \end{figure}
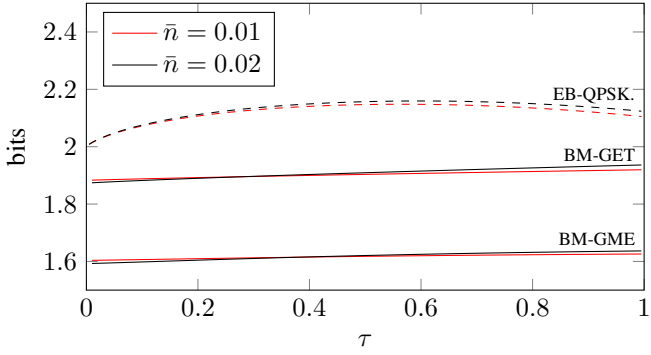

    We compare the three ways of computing Eve entropy: from the EB protocol and from our BM decomposition using either the GET and the GM. In the Appendix \ref{ap:cov-mat} we show the covariance matrix for $\vu\rho$ when Alice apply a QPSK modulation, from which we can evaluate its entropy. In the \Cref{fig:entropy} we plotted the bounds on Eve's entropy with $\alpha = 1$ and for $\bar{n} = \qty{0.01,0.02}$ as a function of the channel transmittance $\tau$. We can see that the entropy values obtained by the EB protocol is a conservative measure and our bound lies bellow on the entire transmittance range. The entropy value computed from the Gramm matrix is even lower, as expected (GET always upper bounds it). Afterwards, one can assume that the expected performance of a discrete modulated CVQKD protocol is more accurate when the eavesdropper entropy is estimated using our model based on the BM decomposition of the after entangling cloner state.
    
    \section{Conclusions}\label{sec:conclusions}

	We presented a method for computing the eavesdropper entropy in the context of discrete modulated CVQKD (Prepare and Measure) protocol with coherent states under the Entangling Cloner Attack. Our method uses the Block-Messiah decomposition to describe the eavesdropper TMSV evolution and we found that her average state, induced by Alice's ensemble, has the same entropy of a set of displaced two-mode thermal states whose displacements are a function of the complex amplitudes of Alice's states. We compared our method with the entropy obtained by using the entangled based version and resulted that our bound on the entropy is lower. In order to compute the Holevo bound on Eve's information, one must compute Eve's entropy conditioned to Bob's outcomes (for reverse reconciliation) and then the covariance matrix conditioned for the set of displaced two-mode thermal states. Future work could also investigate if a set of multimode Gaussian states could have its entropy evaluated by its Gramm Matrix. 
    
    \appendices
    
    \section{The Covariance Matrix of $\vu\rho$}\label{ap:cov-mat}

    Let $AB$ be a two-mode composite system and $\vu\rho_k = \vu{D}_{d_k}\vu\rho_{\Bar{n}}\otimes\vu\rho_{\Bar{m}}\vu{D}_{d_k}^\dagger$ a bipartite displaced thermal state on it with mean photon numbers $\Bar{n}$ and $\Bar{m}$, where $\vu{D}_{\bm{d}_k} = \vu{D}_{x_k}\otimes\vu{D}_{y_k}$ is a two mode displacement operator on $AB$ with $x_k = -xe^{i\theta_k}$, $y_k = ye^{-i\theta_k}$ where $\theta_k = \qty{\pi/4, 3\pi/4, -\pi/4, -3\pi/4}$. If $p_k = 1/4$ is the probability assigned to each state $\rho_k$, the average state $\rho = \frac{1}{4}\sum_k\rho_k$ has the same characteristics of the state in \Cref{eq:thermal-displaced-state}, whose entropy is the same of the eavesdropper state \Cref{eq:eve-state}. Then, we are interested in developing the expressions for the second statistical moment of $\rho$.

    The covariance matrix for a two mode state is a $4\times 4$ real symmetric and has the following standard block form 
    \begin{equation}
        \bm\Sigma = \begin{pmatrix} \bm\Sigma_A & \bm\Sigma_C \\ \bm\Sigma_C^T & \bm\Sigma_B \end{pmatrix},
    \end{equation}
    \noindent where $\bm\Sigma_A$, $\bm\Sigma_B$, $\bm\Sigma_C$ are $2\times 2$ matrices corresponding to modes $A$, $B$ and the correlations between their quadratures, respectively. Its elements are $\Sigma_{jk} = \frac{1}{2}\ev{\qty{\Delta\opr_j}, \qty{\Delta\opr_k}}$, where $\Delta\opr_i = \opr_i - \ev{\opr_i}$ and $\qty{,}$ is the anticomutator.

    Firstly, let us call $\opa_1$ and $\opa_2$ the anihilators for modes $A$ and $B$, respectively. Then, we have the following properties,
    \begin{align}
        \ev*{\opa_i} &= \ev*{\opad_i} = 0, & \ev*{\opad_1\opa_1} &= \Bar{n}+x^2,\\
        \ev*{\opa_i^2} &= \ev*{\opa_i^{\dagger 2}} = 0, & \ev*{\opad_2\opa_2} &= \Bar{m}+y^2,\\
        \ev*{\opa_1\opad_2} &= \ev*{\opad_1\opa_2} = 0, & \ev{\opa_1\opa_2} &= \ev*{\opad_1\opad_2} = -xy,
    \end{align}
    \noindent from which follows that the covariance matrix is
    \begin{equation}
        \bm\Sigma = \begin{pmatrix}
        \qty(2(\Bar{n} + x^2) + 1)\bm{I} & -2xy\bm{Z}\\
        -2xy\bm{Z}                  & \qty(2(\Bar{m} + y^2) + 1)\bm{I}
        \end{pmatrix}.
    \end{equation}

    \bibliographystyle{plain}
    \bibliography{2021-QCQC.bib}
    
\end{document}